\documentclass[aps,pre,twocolumn,groupedaddress,showpacs,showkeys]{revtex4}

\usepackage{graphicx}
\usepackage{dcolumn}
\usepackage{bm}
\usepackage{amssymb}

\begin{document}

\title{Controlling chaos in spatially extended beam-plasma system by the continuous delayed feedback\footnote{\sf Published in \large CHAOS. {\bf16} (2006) 013123}}

\author{Alexander~E.~Hramov}
\email{aeh@nonlin.sgu.ru, aeh@cas.ssu.runnet.ru}
\author{Alexey~A.~Koronovskii}
\email{alkor@nonlin.sgu.ru}
\author{Irene~S.~Rempen}
\email{rempen@cas.ssu.runnet.ru}
\affiliation{Department of
Nonlinear Processes, Saratov State University, Astrakhanskaya, 83,
Saratov, 410012, Russia}

\date{\today}

\begin{abstract}
\noindent
In present paper we discuss the control of complex spatio-temporal dynamics
 in a {spatially extended} non-linear system (fluid model of Pierce diode)
 based on the concepts of controlling chaos in the systems with few degrees of
freedom. A presented method is connected with stabilization of
unstable homogeneous equilibrium state and the unstable
spatio-temporal periodical states analogous to unstable periodic
orbits of chaotic dynamics of the systems with few degrees of
freedom. We show that this method is effective and allows to
achieve desired regular dynamics chosen from a number of possible
in the considered system.
\end{abstract}

\pacs{05.45.-a, 05.45.Gg, 52.35.-g, 52.35.Mw}
\keywords{controlling chaos, spatially extended system,
spatio--temporal chaos, Pierce diode, beam plasma system, unstable
periodic spatio--temporal states}

\maketitle

{\bf The problem of controlling dynamics of non-linear dynamical
systems by stabilization of existing unstable periodical orbits
divides into two tasks. The first of them is picking out and
analysing the periodical states. For this purpose the algorithm of
Lathrop and Kostelich and algorithm of Schmelcher and Diakonos
(SD--method) are often used. The second task is stabilization of
unstable orbit. In distributed systems the role of unstable orbits
is performed by spatio-temporal periodical states.  In our work we
propose methods of picking out unstable periodical spatio-temporal
states with the help of modified Kostelich and SD--methods for
distributed beam-plasma systems. Stabilization is realized with
the help of continuous delayed feedback, by the method based on
Pyragas works. The most attractive feature of the proposed method
is that the continuous control signal is given to one of the
boundaries of the system what makes this method convenient to use
in practice, for example, for microwave beam-plasma systems.}

\section*{Introduction}

Practical use of unstable periodic orbits of chaotic attractor for
chaos controlling has been discussed in a number of papers since
the the work of E.\,Ott, C.\,Grebogy, J.\,Yorke
\cite{Ott:1990_ControllingChaos} has been published. In the works
\cite{Boccaletti:2000_ControlReview, Grebogi:1988_UnstableOrbits,
Kostelich:1989_experiment, Schmelcher:1997_UnstableOrbit,
Dhamala:1999_UnstableOrbits, Carroll:1999_UnstableOrbits,
Gallas:2001_UnstableOrbits, Pingel:2001_SD-methodPRE} it has been
shown that unstable orbits of chaotic attractor of a dynamical
system with few degrees of freedom can be used to control complex
behaviour of the system, and the stabilization of the unstable
periodic orbit needs small perturbation of the system dynamics.
The work \cite{Ott:1990_ControllingChaos} offers an algorithm of
chaos control, defining the change of the system parameter in the
dependence of the distance between the current state of the system
and the wanted orbit. The value and sign of the parameter change
can be discussed as the signal of some feedback. This method can
also be applied to flow systems.

As it has been shown in a number of works
\cite{Ditto:1990_ControlChaos, Hunt:1991_ControllingChaos,
Bielawski:1993_ControllingChaos, Shinbrot:1993_ControlChaos} the
OGY algorithm is insensitive to noise and inexact knowledge of the
system state and it can be applied to the systems with finite, but
large degrees of freedom \cite{Ding:1996_ControllingChaos}.

But it seems rather hard to apply the OGY algorithm to control the
dynamics of different distributed chaotic systems, including one
we are discussing in this work, because it demands the explicit
knowledge of system state and the quick change of control
parameter along the whole system space. {It is nearly impossible
to realize in real microwave beam-plasma systems working in the
frequency rang $0.1\div100$~GHz.}

For such systems it seems more prospective to use the method of
stabilization of the unstable periodic states suggested by
K.\,Pyragas \cite{Pyragas:1992_ControllingChaos}. In this work the
continuous \cite{Chen:1994_ControllingChaos}, and not discrete
quickly changing feedback is used, what makes this algorithm
appropriate for the distributed microwave systems. In the Pyragas
scheme the system is synchronized with its own state taken one
orbit period earlier, by continuous change of control parameter
$\varepsilon (t)=\gamma({\xi(t)-\xi(t-T)})$, where $\xi(t)$ is the
analyzed variable, $\gamma$ is the feedback coefficient and $T$ is
the orbit period. When the stabilization of the unstable orbit
takes place, the feedback signal has the order of the noise level.
It is important that all the information demanded for
stabilization, except for $T$ and $\gamma$, is contained in the
time series of the system dynamics $\xi(t)$, i.e. is defined
automatically in real time. Applications of such controlling
scheme were theoretically and practically investigated in
different systems and models \cite{Boccaletti:2000_ControlReview,
Chen:1994_ControllingChaos, Gauthier:1994_ControlingChaos,
Elmer:1998_ControlChaos, Kouomou:2002_ControllingChaos} including
laser physics \cite{Roy:1992ControlChaos, Meucci:1994ControlChaos,
Meucci:1996ControlChaos}, models of geophysical processes
\cite{Tziperman:1997} and the processes of ``reaction--diffusion''
\cite{Franceschini:1999}.

We must mark especially that this scheme with continuous feedback
has been used for controlling spatio-temporal chaos in distributed
systems (see for example works \cite{Franceschini:1999,
Lu:1996_ControlChaos, Martin:1996_ControlChaos} devoted to
stabilization of two-dimensional structures in the $2D$
distributed chaotic system, described by partial derivative
equations; works \cite{Gang:1994_ControlChaos,
Grigoriev:1997_ControlChaos, Parmananda:1997}, in which chaos
control in the lattices of coupled maps has been considered; the
works analyzing such sample equations as the complex
Ginzburg-Landau equation in non-linear dynamics
\cite{Montagne:1997_ControllingChaos,
Boccaletti:1999_ControllingChaos} or the Swift-Hohenberg equation
of laser physics \cite{Bleich:1997_ControllingChaos,
Hochheiser:1997_ControlChaos}). Among such investigations we must
mark works \cite{Boccaletti:1999_ControllingChaos,
Franceschini:1999} especially. Thus, in the work of Boccaletti
\cite{Boccaletti:1999_ControllingChaos} the problem of chaos
control in the Ginzburg-Landau model with the use of a large (but
finite) number of small local perturbations has been investigated.
This method allows to stabilize the unstable structures of chaotic
spatio-temporal dynamics and to synchronize two chaotic states.
Franceschini and the others in work \cite{Franceschini:1999}
studied the problem of controlling chaotic generation of impulses
in the model of uniformly coupled ``reaction--diffusion'' systems
used for describing the processes of charge transfer in bistable
semiconductor. The controlling feedback signal in mathematical
model offered in work \cite{Franceschini:1999} is supposed to be
spatially homogeneous, i.e. the perturbation carried into every
point of the system space is the same.

Some works describing chaos control in microwave beam--plasma systems must
also be mentioned. The stabilization of unstable
periodic orbits in a fluid model of Pierce diode with the help of
the OGY algorithm has been discussed in work
\cite{Krahnstover:1998_PierceDiode}. Work
\cite{Friedel:1998_ControllingChaosPierceDiode} discusses the
problem of suppression of the regime of the virtual cathode in
Pierce diode \cite{Pierce:1944, Matsumoto:1996} with the help of
delayed feedback using a numerical method of large particles. In
our work \cite{Hramov:2004_IJE} the influence of the external
feedback on the chaotic oscillations in Pierce diode has been
examined.

It is important to note, that in the majority of works devoted to
chaos control in the distributed systems, the controlling feedback
signal effects at every point of the system space. Such
distributed spatio-temporal control can be used, for example, in
optical systems \cite{Mamaev:1998, Pastur:2004,
Arecchi:1993_Laser}. But application of such methods to
beam-plasma systems causes several difficulties. In the present
work we propose a method of chaos control based on the continuous
feedback scheme, where the feedback signal is given to one of the
boundaries of the distributed system. As a sample this method is
applied to widely known simple distributed beam-plasma system
(fluid model of Pierce diode)

As it has already been mentioned above, the knowledge of the
positions of the unstable orbits in the phase space and their periods is
quite important for controlling chaos in the discrete and flow
systems with a small number of degrees of freedom. We can say that in
distributed systems the unstable periodic spatio-temporal states play
the roles of unstable orbits. Therefore, for the efficient chaos
controlling we need to find such unstable periodic states.
Accordingly, we propose a simple method of detecting of unstable
periodic spatio-temporal states and their further stabilization.

The structure of the paper is following: In
Section~\ref{lbl:GenForm} the fluid model of Pierce diode is
discussed. In Section~\ref{lbl:EQuilibriumState} stabilization of
the unstable homogeneous equilibrium state is described. In
Section~\ref{lbl:States} the spatio-temporal data series of
electron beam dynamics are analyzed and the unstable periodic
spatio--temporal states of the chaotic dynamics are carried out.
And, in Section~\ref{lbl:Stabilization} we discuss the
stabilization of the unstable periodic spatio-temporal states
(unstable states) derived in Section~\ref{lbl:States}, with the
help of feedback of different type. In conclusion we summarize the
main results discussed in our paper.

\section{General formalism}
\label{lbl:GenForm}

\begin{figure}
\centerline{\scalebox{0.35}{\includegraphics{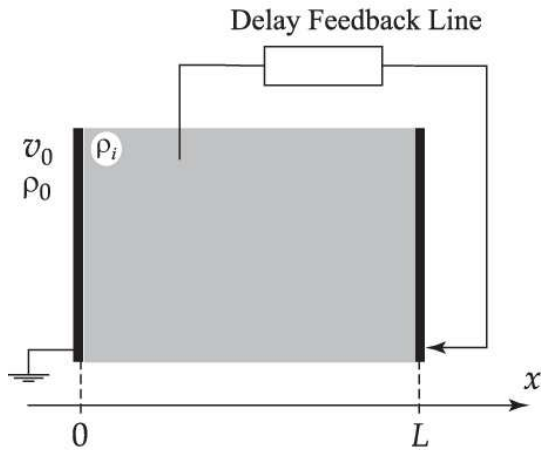}}}
\caption{Schematic diagram of Pierce diode with a feedback line}
\label{ref1}
\end{figure}

Pierce diode \cite{Pierce:1944, Matsumoto:1996} is one of the
simplest beam-plasma systems demonstrating complex chaotic
dynamics \cite{Godfrey:1987, Kuhn:1990, Lindsay:1995,
Matsumoto:1996, Hramov:2004_IJE}. It consists of two plane
parallel infinite grids pierced by the monoenergetic (at the
entrance) electron beam (see Fig.\,\ref{ref1}). The grids are
grounded and the distance between them is $L$. The entrance space
charge density $\rho_0$ and velocity $v_0$ are maintained
constant. The space between the grids is evenly filled by the
neutralizing ions with density $|\rho_i/\rho_0|=1$. The dynamics
of this system is defined by the only parameter, the so-called
Pierce parameter
\begin{equation}\label{q_alpha}
\alpha={\omega_pL}/{v_0},
\end{equation}
where $\omega_p$ is the plasma frequency of the electron beam.
With $\alpha>\pi$ in the system, the so-called Pierce instability
\cite{Pierce:1944, Matsumoto:1996} develops, which leads to the
appearance of the virtual cathode. At the same time, with
$\alpha\sim3\pi$, the instability is limited by non-linearity and
the regime of complete passing of the electron beam through the
diode space can be observed. In this case the system can be
described by the fluid equations:
\begin{equation}\label{q1}
\frac{\partial v}{\partial t}+v\frac{\partial v}{\partial
x}=\frac{\partial\varphi}{\partial x},
\end{equation}
\begin{equation}\label{q2}
\frac{\partial\rho}{\partial t}+ v\frac{\partial \rho}{\partial
x}+\rho\frac{\partial v}{\partial x}=0,
\end{equation}
\begin{equation}\label{q3}
\frac{\partial^2\varphi}{\partial x^2}=\alpha^2(\rho-1),
\end{equation}
with the boundary conditions:
\begin{equation}\label{q4}
v(0,t)=1,\quad \rho(0,t)=1,\quad\varphi(0,t)=\varphi(1,t)=0.
\end{equation}

In equations (\ref{q1})--(\ref{q4}) the non-dimensional variables
(space charge potential $\varphi$, density $\rho$, velocity $v$,
space coordinate $x$ and time $t$) are used. They are related to
the corresponding dimensional variables as follows:
\begin{equation}\label{q6}
\begin{array}{c}
  \varphi '=  ({v^2_0/\eta})\varphi,\quad E'=({v^2_0/L\eta})E,
  \\[3mm]
   \rho' =\rho_0\rho,\quad v'=v_0 v,\quad x'=L x,\quad t'=({L/v_0})t, \\
\end{array}
\end{equation}
where the dotted symbols correspond to the dimensional values,
$\eta$ is the specific electron charge, $v_0$ and $\rho_0$ are the
non-perturbed velocity and density of the electron beam, $L$ is
the length of the diode space.

Equations (\ref{q1}) and (\ref{q2}) are numerically integrated
with the help of one-step explicit two-level scheme with upstream
differences \cite{Rouch:1976_FluidNumericalBook,
Potter:1973_BookCompPhysics, Matsumoto:1996} and Poisson equation
(\ref{q3}) is solved by the method of the error vector propagation
\cite{Rouch:1976_FluidNumericalBook}. The time and space
integration steps have been taken as $\Delta x=0.0005$, and
$\Delta t=0.0001$, respectively.

In works \cite{Godfrey:1987, Kuhn:1990, Matsumoto:1996} it was
shown, that in the narrow range $\alpha\sim3\pi$ without the
virtual cathode, one can observe chaotic oscillations in the
system. With a decrease of $\alpha$ from $2.88\pi$ to $2.86\pi$
the model demonstrates a transition from periodic oscillations to
chaos via a period doubling cascade. Just below the critical value
of $\alpha$ the system demonstrates weak chaotic oscillations with
the explicit time scale. Below we will call this regime ``bond
chaos'' because the attractor in the reconstructed phase space
looks like a narrow bond on which the phase paths are situated.
With the further decrease of $\alpha$ the explicit time scale
vanishes and the phase picture reminds of spiral twisting from one
point. The power spectrum of the system oscillations is more
complex in comparison with the first case. This type of chaotic
behaviour we call ``spiral chaos''.

The stationary homogeneous equilibrium state of the electron beam
in Pierce diode is characterized by the following distribution of
the system variables
\begin{equation}\label{q18}
 \bar v(x)=1.0,\quad \bar\rho(x)=1.0,\quad \bar\varphi(x)=0.0,
\end{equation}
that can be found from solutions of stationary equations derived
from non-stationary fluid equations (\ref{q1})--(\ref{q3}) in
assumption that $\partial/\partial t=0$.

Homogeneous equilibrium state (\ref{q18}) is stable with
$\alpha<\pi$, and loses its stability at $\alpha>\pi$ as it has
been shown by J.\,Pierce \cite{Pierce:1944}. In work
\cite{Hramov:2004_IJE} the significance of the unstable a decrease
of $\alpha$ or with the influence of the external delayed feedback
is shown.

\section{Stabilization of unstable equilibrium {spatial}
state with the help of the continuous delayed feedback}
\label{lbl:EQuilibriumState}

Following the works mentioned above
\cite{Chen:1994_ControllingChaos, Gauthier:1994_ControlingChaos,
Elmer:1998_ControlChaos, Kouomou:2002_ControllingChaos,
Roy:1992ControlChaos, Meucci:1994ControlChaos,
Meucci:1996ControlChaos,Tziperman:1997, Franceschini:1999} devoted
to chaos control in finite-dimensional systems, we now describe
the scheme of the continuous delayed feedback for stabilization of
the unstable equilibrium state in Pierce diode.

The continuous delayed feedback is realized by changing the
potential of the right boundary of the system:
\begin{equation}\label{q_OS}
 \varphi(x=1.0,t)=f_{\rm fb}(t)=K(\rho(x_{\rm fix},t)-\rho(x_{\rm fix},t-d)),
\end{equation}
where $K$ is the feedback coefficient, $d$ is the delay time.
Variable $\rho(x_{\rm fix},t)$ in (\ref{q_OS}) is the space charge
density in a fixed point $x=x_{\rm fix}$ of diode space (in the
work we use $x_{\rm fix}=0.2$). When the stabilization regime
takes place and the system is exactly in the equilibrium state,
the feedback signal $f_{\rm fb}$ could by compared with the noise
level.

Practically, this scheme of delaying feedback can be realized, for
example, with the help of delay lines on magnetostatic waves
\cite{Gastera:1984_MSW_devices, Sethares:1982_MSW} or acoustic
waves \cite{Losee:1997_AcousticDelayLine,
Filipiak:1986_AcousticDelayLine}. Such method would allow to
select the required delay time. The signal can be picked out from
the system using a probe.

So, unlike works \cite{Franceschini:1999, Lu:1996_ControlChaos,
Martin:1996_ControlChaos, Gang:1994_ControlChaos,
Grigoriev:1997_ControlChaos, Parmananda:1997,
Montagne:1997_ControllingChaos, Bleich:1997_ControllingChaos,
Hochheiser:1997_ControlChaos}, where controlling the chaotic
dynamics of the distributed systems needs the effect of the
feedback signal upon the whole interaction interval, in our scheme
the signal of the continuous delayed feedback determines only the
change of the boundary conditions. It is more convenient to
realize in experimental microwave devices. Then, in real microwave
systems working in the frequency range $0.1\div100$~GHz, it is
nearly impossible to realize the OGY algorithm because it demands
very quick changing of controlling parameter.

It must be marked that we use the term continuous feedback in
analogy with works \cite{Pyragas:1992_ControllingChaos,
Chen:1994_ControllingChaos}. This term emphasizes that in our
scheme the controlling signal is changed continuously in contrast
to the schemes based on the OGY algorithm
\cite{Ott:1990_ControllingChaos}. It is obvious that the
controlling scheme~(\ref{q_OS}) discussed in our work is more
convenient in practical use for controlling the dynamics of the
distributed microwave systems.

For purpose of stabilization of the unstable equilibrium state the
delay time $d$ of the feedback line (\ref{q_OS}) must be small
enough:
 $d<T_p/2$, where $T_p=\pi/\omega_p$ is the typical time scale in
 the autonomous distributed system \cite{Matsumoto:1996}.

\begin{figure}[t]
\centerline{\scalebox{0.4}{\includegraphics{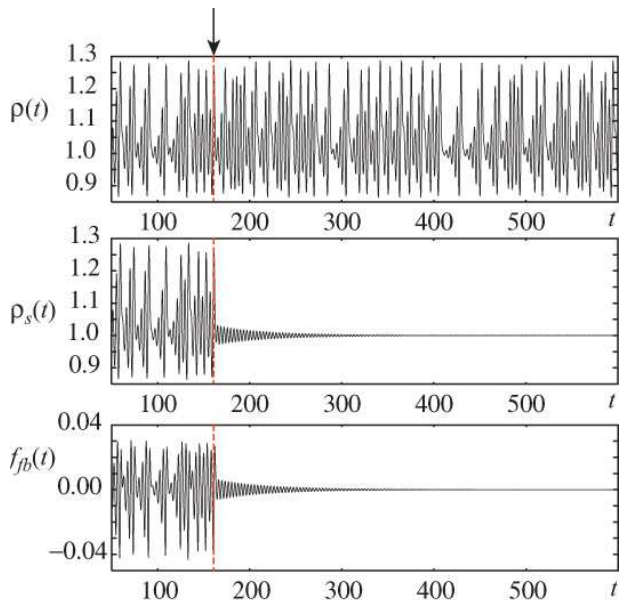}}}
\caption{Time series $\rho(x_{\rm fix},t)$ in autonomous system
 (top picture), oscillations $\rho_s(x_{\rm fix},t)$ in stabilized system
(middle picture) and the signal in feedback line $f_{\rm fb}(t)$
(bottom picture) for the spiral chaos regime ($\alpha=2.857\pi$).
The feedback parameters are $K=0.8$, $d=0.15$. The moment of
appearance of the delayed feedback is marked by an arrow and
dotted line.} \label{ref8}
\end{figure}

Results of stabilization of the system (\ref{q1})--(\ref{q4}) with
the help of the continuous feedback (\ref{q_OS}) are presented in
Fig.\,\ref{ref8}. The autonomous system displays chaotic
oscillations with a large amplitude. But after switching on the
continuous feedback one can observe a fast decrease of the
amplitude of oscillations which leads to the stabilization of the
unstable state (\ref{q18}). After a short transient process the
controlling signal in the feedback line becomes rather small in
comparison with the signal before stabilization (near 0.01\%).
This means that the regime of chaos control with the help of small
controlling signal of continuous feedback takes place in the
system.

\begin{figure}[t]
\centerline{\scalebox{0.4}{\includegraphics{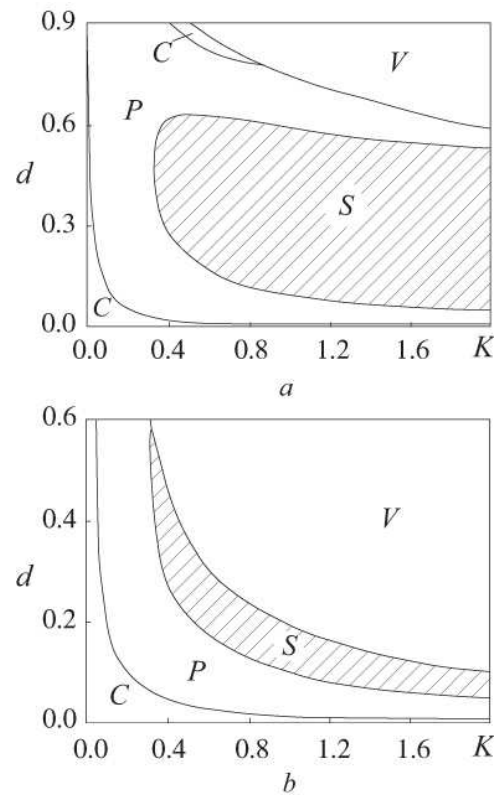}}}
\caption{Regime maps of the stabilized system on the plane of
parameters $(K, d)$ for cases of ({\it a}) bond chaos
($\alpha=2.862\pi$) and ({\it b}) spiral chaos
($\alpha=2.857\pi$). The region of stabilization of the unstable
equilibrium state is hatched} \label{ref9}
\end{figure}

It is important to find the regions of feedback parameters $K$ and
$d$ where chaos controlling is possible. Fig.\,\ref{ref9} shows
areas of the bond chaos regime ($\alpha=2.862\pi$) and spiral
chaos regime ($\alpha=2.857\pi$). One can see that the regime maps
($K,d$) for both cases have qualitative similarities. For small
$K$, the system displays chaotic oscillations similar to
autonomous oscillations corresponding to this $\alpha$ value
(region $C$ in Fig.\,\ref{ref9}). With an increase of $K$ one can
observe the destruction of chaotic oscillations and appearance of
periodical ones (region $P$). In this case the feedback signal is
not small (its amplitude has the same order of value as $f_{\rm
fb}$ before switching on the feedback) hence we can not regard
this regime as the controlling chaos one. With further growth of
$K$ the controlling chaos regime takes place (region $S$). The
unstable homogeneous state is stabilized, and the system dynamics
in this case is shown in Fig.\,\ref{ref8}.

The width of region of stabilization $S$ depends noticeably on the
delay time $d$. The threshold values $d_1$, $d_2$, determine the
range $d\in(d_1,d_2)$ where controlling parameter values allowing
to obtain controlling chaos in the distributed beam-plasma system
are possible.

With the further increase of $d$ and $K$, reflection and
overtaking of electrons appear in the electron beam (region $V$ in
Fig.\,\ref{ref9}), and therefore the fluid equations
(\ref{q1})--(\ref{q4}) become unfit (see
\cite{Shevchik:1966_Microwave,
Soohoo:1971_MicrowaveElectronicsBook} for detail).

It is useful to note that in spiral chaos regime the region $S$ in
the parameter map is narrower than that in bond chaos regime
(compare Fig.\,\ref{ref9}{\it a} and~\ref{ref9}{\it b}).

We can analyse the stability of the discussed equilibrium state by
calculating the maximum Lyapunov exponent $\Lambda$
\cite{Pyragas:1992_ControllingChaos}. For the feedback coefficient
$K=0$ the values of Lyapunov exponents of our system are equal to
those of the autonomous system.

\begin{figure}[t]
\centerline{\scalebox{0.4}{\includegraphics{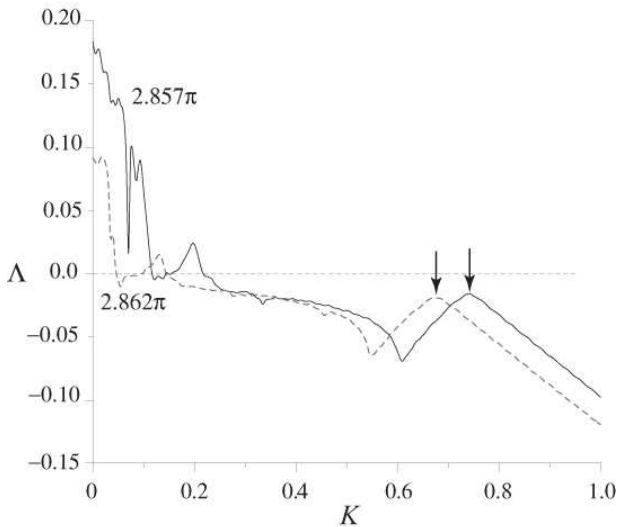}}}
\caption{Dependence of the maximum Lyapunov exponent $\Lambda$
upon the feedback coefficient $K$. The dotted line corresponds to
the bond chaos regime ($\alpha=2.862\pi$), the solid line --- to
the spiral chaos regime ($\alpha=2.857\pi$). The arrow marks the
values of $K_c$, corresponding to the onset of stabilization.
$d=0.14$} \label{ref10}
\end{figure}

Fig.\,\ref{ref10} shows the dependence between the maximum
Lyapunov exponent $\Lambda$ and the feedback coefficient $K$ for
two different chaotic regimes. This dependencies determine the
boundary of adaptability of our control method. The suppression of
the chaotic dynamics is possible only for the values of $K$
corresponding to $\Lambda(K)<0$. These are regions $P$ and $S$ in
Fig.\,\ref{ref9}. The value of $K$ for which stabilization of
unstable state takes place (region $S$ in Fig.\,\ref{ref9}) is
marked by an arrow in Fig.\,\ref{ref10}. One can see that the
stabilization regime is preceded by the increase of Lyapunov
exponent, and beyond the boundary of stabilization $K=K_c$ the
value $\Lambda(K)$ decreases linearly. Such behaviour is typical
to both types of chaotic dynamics.

\section{Picking out unstable periodical spatio-temporal states of {spatially extended system} dynamics}
\label{lbl:States}

In this section we discuss the unstable periodic spatio-temporal
states of distributed active medium, analogous to the unstable
periodic orbits of a few-dimensional systems, which play an
important role in the complex behavior of non-linear systems
\cite{Cvitanovic:1991_orbits}, in particular, in the case of
chaotic synchronization \cite{Boccaletti:2002_ChaosSynchro,
Mendoza:2004, Leyva:2003_SynchroChaos,
Pikovsky:1997_PhaseSynchro_UPOs, Pazo:2002_UPOsSynchro,
Aeh:2005_SpectralComponents}.

The initial information about the set of unstable periodic orbits
can be obtained from the histograms describing the frequency of
system returning to the vicinity of the orbits. This method was
offered by  D.P.\,Lathrop и E.J.\,Kostelich
\cite{Kostelich:1989_experiment}. We choose a phase point ${\bf
R}_i$ on the attractor. If this point is close to an unstable
cycle with a period $T$, then the phase trajectory passing the
points ${\bf R}_{i+1}$, ${\bf R}_{i+2}$, \dots, ${\bf R}_{k}$,
will come near the initial state with some precision
$\varepsilon>0$:
\begin{equation}\label{q19}
 ||{\bf R}_i-{\bf R}_{i+m}||<\varepsilon,
\end{equation}
where $m=T/ \Delta t$ is the orbit period in discrete time unit.
Then the distribution of return time (histogram) is made. By this
histogram it is easy to carry out the typical periods of
corresponding unstable states, and then to find the unstable
cycles.

For the study of unstable spatio-temporal states of the dynamics
of the explored distributed system we investigate the time
oscillations of the space charge density $\rho(x_0,t)$, obtained
from the fixed points of the interaction space $x_0$. Then, from
the time series $\rho(x_0,t)$, using Takens' method
\cite{Takens:1981} the attractors of system dynamics are restored
in pseudo-phase space $R^n$. Using Lathrop \& Kostelich algorithm
we pick out the periodical orbits in $R^n$ and then hence find the
spatio-temporal periodical states of the distributed system.

\begin{figure}[t]
\centerline{\scalebox{0.4}{\includegraphics{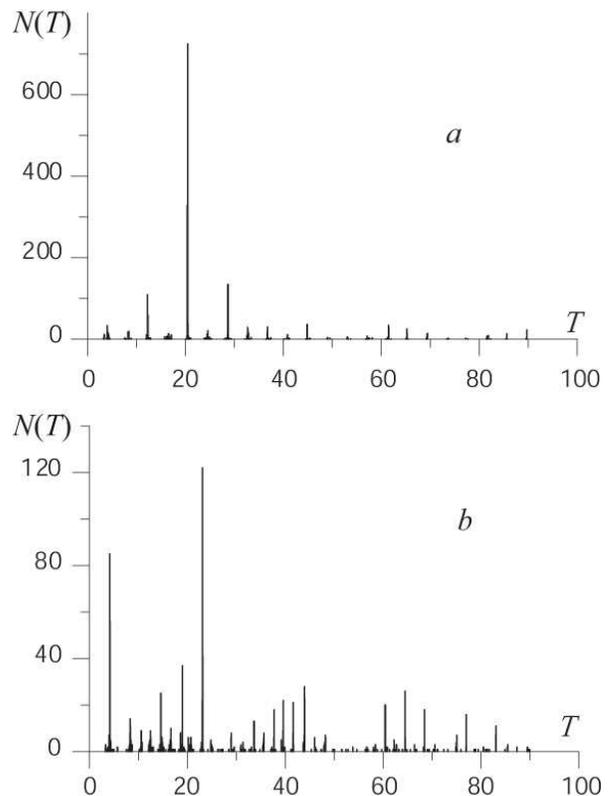}}}
\caption{Histograms showing the number of visits of phase
trajectory to the fixed attractor points, associated with
different recurrence time: ({\it a}) $\alpha=2.861\pi$; ({\it b})
 $\alpha=2.857\pi$}
 \label{ref2}
\end{figure}

In Fig.~\ref{ref2} the histograms of return time obtained for
different values of Pierce parameter are presented. One can see
that in the bond chaos regime (Fig.~\ref{ref2}{\it а}) one of the
orbits dominate upon the others i.e. it is more frequently visited
by phase point.

For spiral chaos regime (Fig.~\ref{ref2}{\it b}) the set of
unstable orbits is more complex, and the different unstable
periodical states are visited more evenly and frequently.

\begin{figure}[t]
\centerline{\scalebox{0.65}{\includegraphics{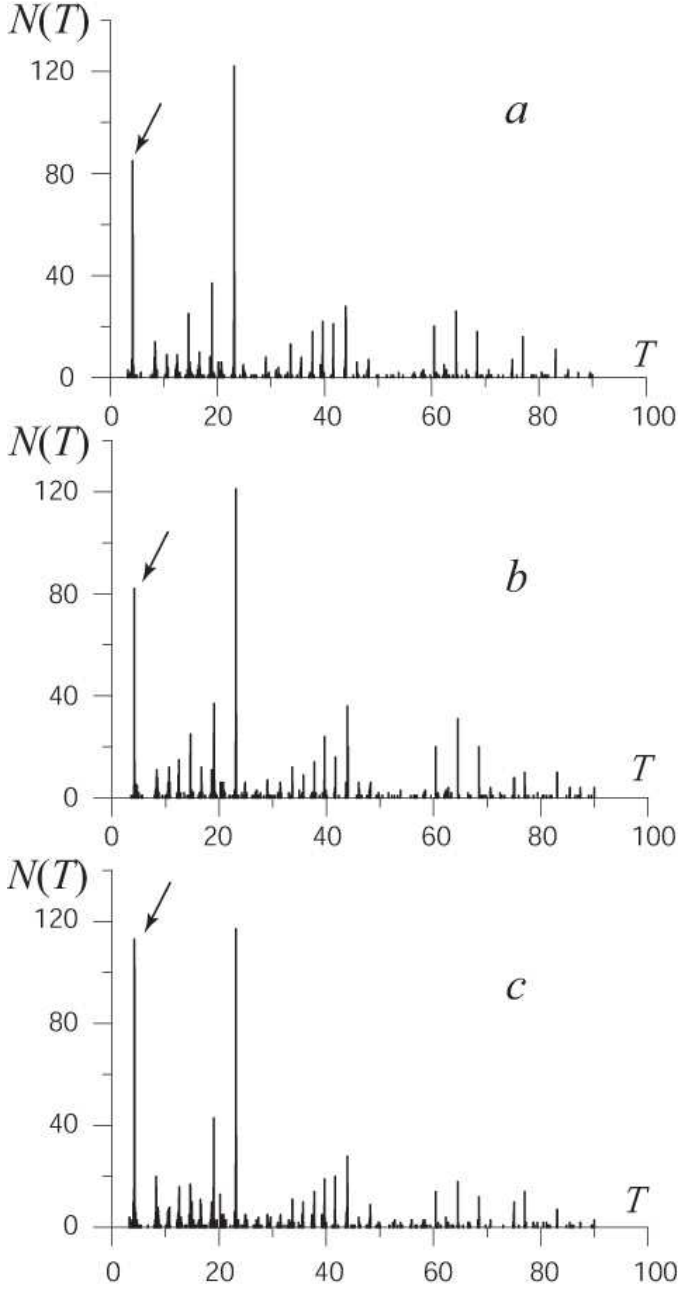}}}
\caption{Histograms of recurrence time for $\alpha=2.857\pi$
obtained from the data taken from different points of diode space:
({\it a}) $x_0=0.2$, ({\it b}) $x_0=0.5$, ({\it c}) $x=0.8$}
 \label{ref3}
\end{figure}

Histograms of recurrence time obtained for time series
$\rho(x,t)$, taken from different points of diode space $x=x_0$
are very much alike. This is illustrated in Fig.~\ref{ref3}. It
follows from the picture that the set of unstable spatio--temporal
states is identical along the diode interval. A slight difference
can be caused by inaccuracy in calculating the recurrence time
(see, for example, the unstable spatio--temporal states with
$T=4.173$ in Fig.~\ref{ref3}{\it a} and Fig.~\ref{ref3}{\it c}
(marked by an arrow)). The latter allows to carry out and to
analyse some characteristics of the unstable spatio-temporal
states of the distributed system using scalar time series obtained
from one fixed point $x_{fix}$ of interaction interval.

\begin{figure}[t]
\centerline{\scalebox{0.4}{\includegraphics{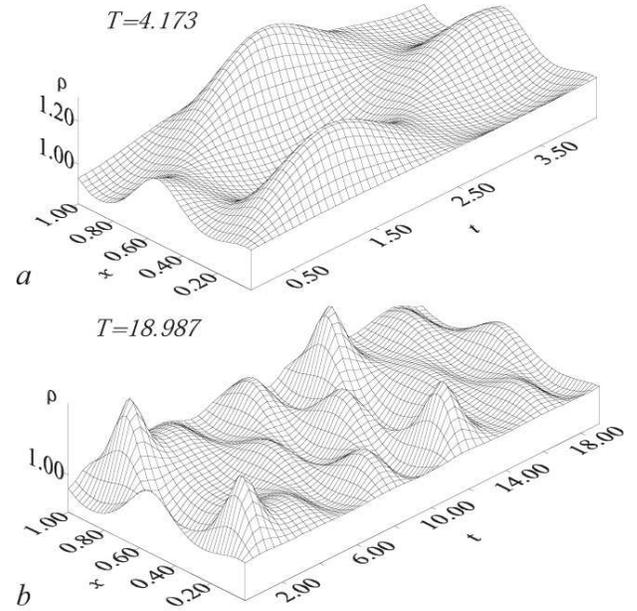}}}
\caption{Unstable periodical orbits of different periods $T$ in
the regime of spiral chaos: ({\it a}) $T=4.173$ and ({\it b})
$T=18.987$}
 \label{ref5}
\end{figure}

In Fig.\,\ref{ref5} the unstable periodical states of the
distributed system are shown. The three-dimensional pictures show
the spatio-temporal oscillations of the space charge density
$\rho(x,t)$ which correspond to the unstable periodical orbit
states with different period. The pictures are obtained for
$\alpha=2.857\pi$ (spiral chaos).

We must pay attention that the similar results has been derived
when the unstable periodic spatio-temporal states by the method of
P.\,Schmelcher и F.\,Diakonos (SD--method)
\cite{Schmelcher:1997_UnstableOrbit, Pingel:2001_SD-methodPRE},
adapted for the analysis of distributed system. The algorithm is
the follows.

As in the case when the unstable periodic states are reconstructed
from histograms, at first we determine the system state vector ${\bf R}$
in phase space. As a state vector in this case a vector is
taken combined from the density of charge taken from the different
points of the interaction space: ${\bf
R}(t)=\left\{{\rho(x=0.25,t),\rho(x=0.5,t),\rho(x=0.75,t)}\right\}^T$.

In reconstructed phase space a plane $\rho(x=0.25,t)=1.0$ is
taken, which is considered as Poincare section. Let us indicate as
${\bf R}_n$ the system state corresponding to the $n$th crossing
between phase part and the selected section. (this crossing takes
place at moment of time $t^{n}$). Then the description of the
system can be made with the help of discrete map like
\begin{equation}\label{q190}
 {\bf R}_{n+1}={\bf G}({\bf R}_{n}),
\end{equation}
where ${\bf G}(\cdot)$ is the evolution operator. Obviously, it is impossible to
find the analytical form for the operator $\bf G$, but numerical integration of the initial system of
fluid equations can give us a consecution of values $\left\{{\bf
R}\right\}_{n}$, derived from the map (\ref{q190}).

SD--method for picking out unstable periodic orbits supposes
examination of modified model of a distributed system, described
by following map \cite{Pingel:2001_SD-methodPRE}:
\begin{equation}\label{q191}
 {\bf R}_{n+1}={\bf R}_{n}+\lambda {\bf C}\left[{\bf G}({\bf R}_{n})-{\bf
 R}_{n}\right].
\end{equation}
In the works \cite{Schmelcher:1997_UnstableOrbit,
Pingel:2001_SD-methodPRE} it is exactly shown that such modified
system in the case of analysis of discrete systems and the systems
with few degrees of freedom allows to stabilize effectively the
unstable periodical orbits of the initial system, which in the
case of modified system transform from saddle into stable ones
(\ref{q191}). Application of the SD--method to the analysis of
distributed system is also effective for picking out unstable
periodic spatio-temporal states.

In the equation (\ref{q191}) $\lambda=0.1$ is the method constant
and $\bf C$ is a matrix, which, following to work
\cite{Pingel:2001_SD-methodPRE} was taken as
\begin{equation}\label{q192}
 {\bf C}= \left(\begin{array}{cc}
   1 & 0 \\
   0 & 1 \\
 \end{array}\right).
\end{equation}

Modified map (\ref{q191}) allows only to define unstable periodic
state of lowest period $T_1$, which agrees with the single
crossing of the selected Poincare section during time $T_1$. To
find the unstable states with higher period $T_p$, when the phase
path crosses the Poincare section for $p$ times during the period
$T_p$, it is necessary to take a modified map like
\begin{equation}\label{q193}
 {\bf R}_{n+1}={\bf R}_{n}+\lambda {\bf C}\left[{\bf G}^p({\bf R}_{n})-{\bf
 R}_{n}\right],
\end{equation}
where ${\bf G}^p(\cdot)$ is $p$ times iterated map ${\bf
G}(\cdot)$ (i.e. when numerical solution of the system of fluid
equations is found and then the map is reconstructed (\ref{q193})
it is necessary to take into account only the every $p$th crossing
of the Poincare section by phase path, where $p$ is obviously integer.

So, by numerical integration of the map (\ref{q193}) with different values of
$p$ one can find the unstable spatio-temporal states that appear to be stable
in the modified system, described by the map (\ref{q193}). So, the result of
numerical integration of modified system (fluid equations taking into account the
procedure (\ref{q193}))) is a periodic solution, which agrees with the
unstable periodic state of the initial system.

The numerical integration of the modified system causes only one
difficulty. When we are searching for the state ${\bf R}_{n+1}$ at
the moment $t^{n+1}$, then, according to formula (\ref{q193}) we
know only the coordinates of this state in Poincare section but we
don't know the corresponding distribution of space charge density
$\rho(x,t^{n+1})$, velocity $v(x,t^{n+1})$ of the electron beam
and the potential $\phi(x,t^{n+1})$. To derive the mentioned space
functions we use the following procedure. The system of fluid
equations describing Pierce diode is integrated till some state
${\bf R}_s$ of the system would not be close to the selected state
 ${\bf R}_{n+1}$  with some demanded precision:
$||{\bf R}_{n+1}-{\bf R}_s||<\delta$, where $\delta$ is taken as
$\delta=10^{-3}$. When this condition is fulfilled, the space
functions corresponding to the state ${\bf R}_s$ are considered as
the space distributions ${\bf R}_{n+1}$ and then the next
iteration according to (\ref{q193}) takes place.

\begin{figure}
\centerline{\scalebox{0.4}{\includegraphics{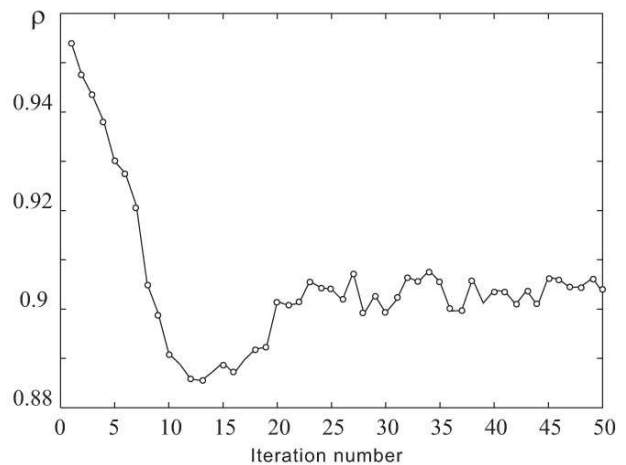}}}
\caption{Dependance of the space charge density $\rho$ taken at
the moment when the phase point passes the Poincare section, upon
the number of iteration of SD--method when the orbit of period 1
($T=4.2$) is picked out. Pierce parameter $\alpha=2.858\pi$}
 \label{Ref:G}
\end{figure}

\begin{figure}
\centerline{\scalebox{0.4}{\includegraphics{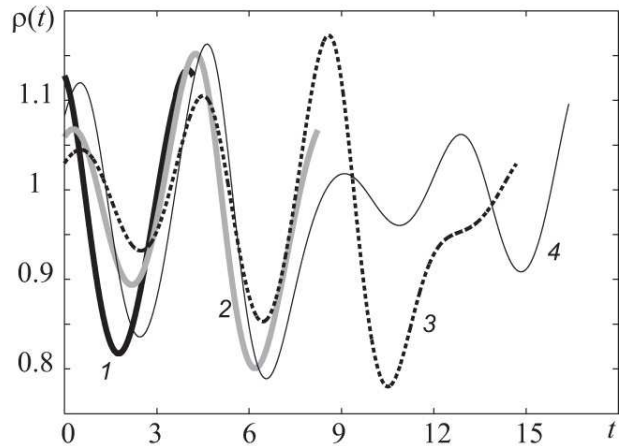}}}
\caption{Time series of the space charge density $\rho(t)$ in the
fixed point of the phase space $x=0.2$, corresponding to the
different unstable periodic spatio-temporal states picked out by
SD--method with $\alpha=2.858\pi$: {\sl1} -- $T=4.2$ ($p=1$),
{\sl2} -- $T=9.1$ ($p=2$), {\sl3} -- $T=16.9$ ($p=3$), {\sl1} --
$T=18.9$ ($p=4$)}
 \label{Ref:Orbits}
\end{figure}

As numerical modelling shows, the modified SD--method  applied to
distributed system is convergent and allows to find the demanded
periodical time-space states. The convergence of the procedure is
illustrated by Fig.~\ref{Ref:G}, which shows the dependance of the
space charge density $\rho(x=0.25)$ in the moment when the  point
in pseudo-phase space crosses the Poincare section upon the number
of iteration $n$ of the SD--method when the unstable state of
period 1 is defined ($p=1$). One can see clearly that the
iteration process of SD--method converges to the value
corresponding to the unstable time-periodical spatio-temporal
state of the system. Fig.~\ref{Ref:Orbits} shows the time series
of $\rho(t,x=0.2)$, corresponding to unstable periodic states of
different order $p$ and period $T$ with $\alpha=2.858\pi$, picked
by SD--method. We mark that the derived spatio-temporal
distributions agrees very well with the distributions of unstable
states derived earlier by analysing of the return time histograms.

As quantitative characteristics of the derived unstable orbits it
is useful to {analyze} the maximum Lyapunov exponent $\Lambda$ of
every orbit. Calculation of this value is important for further
stabilization of periodical spatio-temporal states.

Values of maximum Lyapunov exponent $\Lambda^T$ calculated with
the help of Benettin's algorithm \cite{Benettin:1976} for the most
frequently visited orbits are presented in Table\,1.

\begin{table}
\noindent
  Table. 1: Values of maximum Lyapunov exponent $\Lambda^T$ for
  unstable states with period $T$ in the regime of spiral chaos ($\alpha=2.857\pi$)

\centering

\bigskip

\begin{tabular}{cc}
  \hline
  Period of unstable & Maximum Lyapunov \\
  state $T$ & exponent $\Lambda^T$ \\
  \hline
  \hline
  4.347 & 0.854 \\
  8.289 & 1.194 \\
  10.581 & 0.698 \\
  12.501 & 0.629 \\
  14.613 & 0.337 \\
  18.987 & 0.266 \\
  23.115 & 0.186 \\
  39.636 & 0.094 \\
  43.953 & 0.083 \\
  60.438 & 0.053 \\
  \hline
\end{tabular}
\end{table}

\section{Stabilization of the unstable periodical states}
\label{lbl:Stabilization}

As simplest scheme of stabilization of unstable periodical orbits
described in the previous section, we can take the scheme
(\ref{q_OS}) where the feedback signal is formed as:
$$
 \varphi(x=1.0,t)=f^T_{\rm fb}(t)=
$$
\begin{equation}\label{q_OSS}
=K(\rho(x_{\rm fix},t)-\rho(x_{\rm fix},t-T_k))=K\xi(t).
\end{equation}
Here $K$ is again the feedback coefficient and $T_k$ is the delay
time equal to the period of the $k$-th unstable orbit. As in the
case of stabilization of unstable equilibrium state, we choose the
fixed point of the diode space $x_{\rm fix}=0.2$, from which the
feedback signal is taken.

\begin{figure}
\centerline{\scalebox{0.4}{\includegraphics{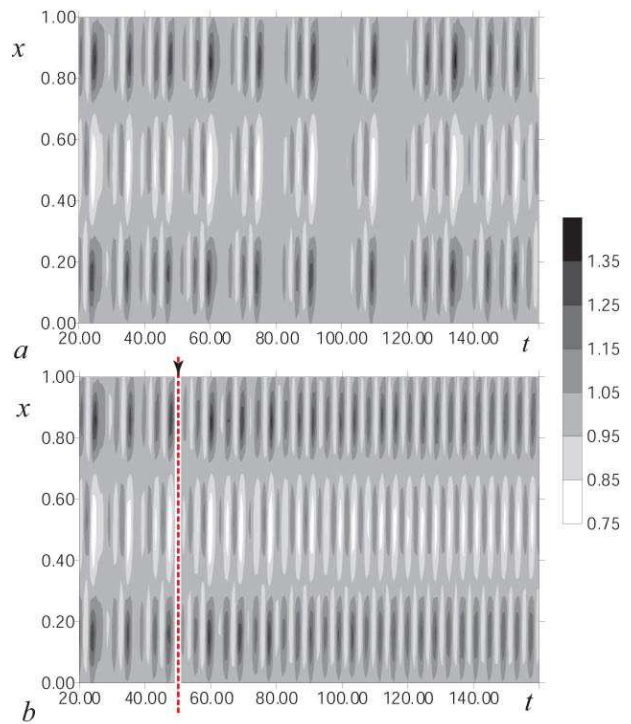}}}
\caption{Spatio--temporal dynamics of Pierce diode in chaotic
regime ({\it a}) and in the regime of stabilization of the
periodic orbit $T=4.173$ ({\it b}). The arrow and dotted line in
\,{\it b} shows the moment of switching on the delayed feedback}
 \label{ref12}
\end{figure}

The numerical modelling shows that this scheme is rather effective
for stabilization of the unstable periodical state with the lowest
period $T_1$. Fig.~\ref{ref12} shows the spatio--temporal dynamics
of the system in cases of autonomous oscillations and in the
regime of stabilization (the value of the space charge density
$\rho(x,t)$ is shown by colour nuances scaling). One can easily
observe that there are peculiarities of the complex
spatio-temporal behaviour and transitions between chaotic and
periodic regimes. It is obvious from the picture that the periodic
dynamics analogous to the system behaviour near the unstable state
is stabilized during approximately $2\div3$ time periods $T_1$
(see Fig.\,\ref{ref5}{\it a}).

\begin{figure}
\centerline{\scalebox{0.4}{\includegraphics{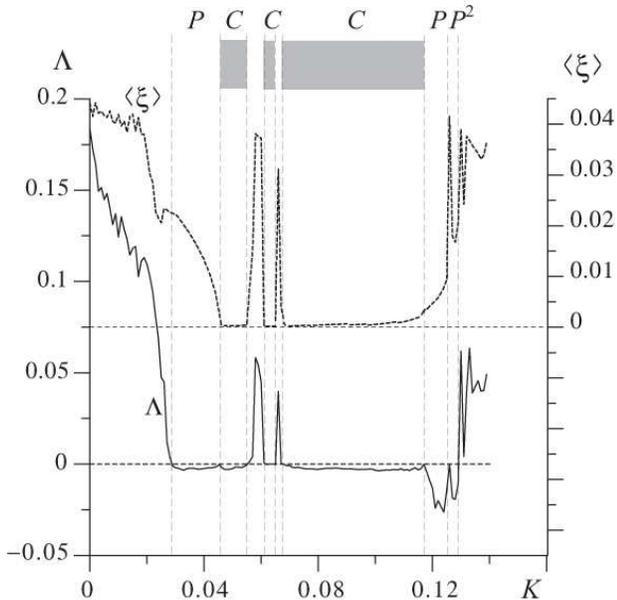}}}
\caption{Dependence of the maximum Lyapunov exponent (solid line)
and the mean feedback signal (doted line) upon the feedback
coefficient $K$ in the spiral chaos regime ($\alpha=2.857\pi$).
Region $C$ (grey colour) corresponds to the regime of
stabilization of the unstable periodic state of period 1 $T=4.173$
(the feedback signal is close to null
$\langle{\xi}\rangle\approx0$). Region $P$ corresponds to periodic
oscillations, when the feedback signal is large and the maximum
Lyapunov exponent $\Lambda<0$; region $P^2$ corresponds to the
regime of period doubling}
 \label{ref13}
\end{figure}

We also study the peculiarities of the influence of delayed
feedback (\ref{q_OSS}) upon the investigated system by examining
how the feedback amplitude $K$ acts upon the maximum Lyapunov
exponent $\Lambda$ and the average of the feedback signal
\begin{equation}\label{q_aver}
\langle{\xi}\rangle=\frac{1}{\tau}\int_0^{\tau}f^T_{\rm
fb}(t)\,dt,
\end{equation}
where $\tau\gg d$. The corresponding plots derived for $d=T_1$ are
presented in Fig.~\ref{ref13}.

\begin{figure}
\centerline{\scalebox{0.5}{\includegraphics{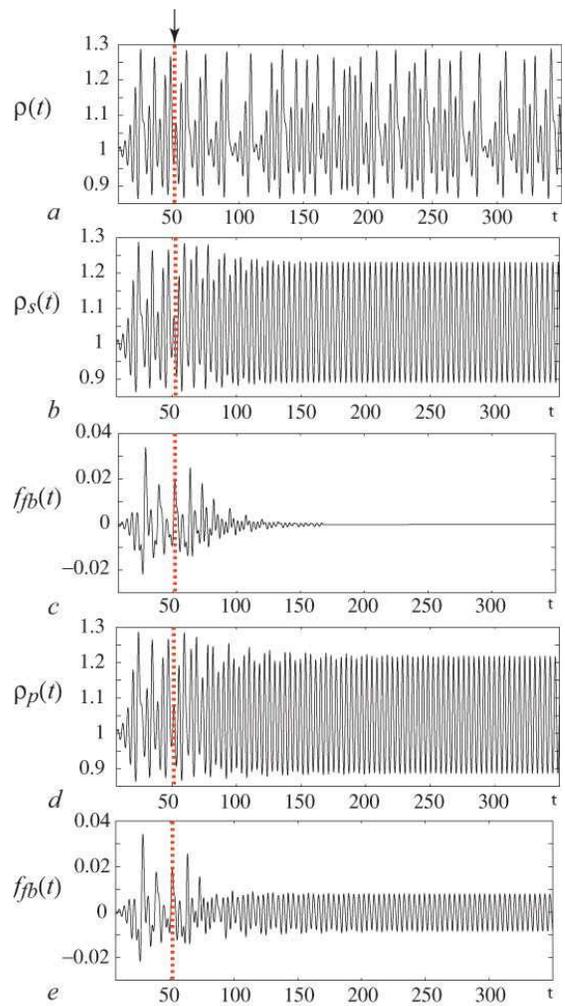}}}
\caption{Time series $\rho(x_{\rm fix},t)$ of the autonomous
system ({\it a}), time series $\rho_s(x_{\rm fix},t)$ in the
stabilized system ($K=0.07$, $d=T=4.173$) ({\it b}), feedback
signal $f_{\rm fb}(t)$ ({\it c}); time series $\rho_p(x_{\rm
fix},t)$ and $f_{\rm fb}(t)$ ({\it d},{\it e}) in periodic
oscillation regime ($\Lambda<0$) ($K=0.035$, $d=T=4.173$),
$\alpha=2.857\pi$ (spiral chaos regime). The moment of switching
on the feedback is marked by an arrow and a dotted line.}
 \label{ref14}
\end{figure}

When $K$ is small the system demonstrates chaotic oscillations
similar to the oscillations in the autonomous regime (in
Fig.~\ref{ref14}{\it a}, one can see the time series of the space
charge oscillations for $x=0.2$ without feedback). With an
increase of $K$ the complexity of the oscillations diminishes (the
value of $\Lambda$ decreases) and at the same time the amplitude
of the feedback signal becomes smaller. In some region of $K$ the
feedback signal is close to zero and the maximum Lyapunov exponent
$\Lambda<0$ (the grey area in Fig.~\ref{ref13}). In this case the
spatio-temporal dynamics is the same as that of the unstable
periodic state. Thus, it is the regime of chaos stabilization,
illustrated by Fig.\,\ref{ref14}{\it b,c}, which shows the
oscillations of the space charge density in the system and the
feedback signal, correspondingly.

Examining the graph $\Lambda(K)$ we also must note the regions
where the feedback signal $\langle{\xi}\rangle$ is not small but
the value $\Lambda<0$. These regimes are marked in
Fig.~\ref{ref13} as $P$ and correspond to periodic oscillations
near the unstable orbit. The space charge oscillations and the
feedback signal for this case are presented in
Fig.\,\ref{ref14}{\it d,e}. We also pay attention to the fact,
that with large $K$ the system demonstrates the period doubling
based on the regime $P$ (region marked as $P^2$ in
Fig.\,\ref{ref13}). For $K>0.15$ in the system the growth of the
oscillation amplitude takes place, and reflection of electrons
appears. Thus, in this region we can not use any more the fluid
equations for the description of the system
\cite{Shevchik:1966_Microwave,
Soohoo:1971_MicrowaveElectronicsBook}.

But the computer experiment shows that stabilization of the
unstable orbits with higher periods $T_k>T_1=4.173$ using feedback
scheme (\ref{q_OSS}) is impossible. The analysis of the chaos
controlling schemes like (\ref{q_OSS}) applied to the systems with
few degrees of freedom, has shown \cite{Just:1999_book} that the
effective stabilization of the unstable periodic orbits is
possible only when the values of the maximum Lyapunov exponent
$\lambda$ and the orbit period $\tau$ fulfill the following
condition
\begin{equation}\label{q_50}
 \lambda\tau\leq C,
\end{equation}
where $C$ is constant depending on the system.

Thus, we must modify the controlling scheme (\ref{q_OSS}) so that
it would be possible to stabilize the orbits with higher period
$T_k$. The scheme, which we consider, is a modification of the
method worked out by K.\,Pyragas
\cite{Pyragas:1992_ControllingChaos}. As it follows from the
result of the numerical experiment
\cite{Socolar:1994_ControllingChaos}, in the few-dimensional
non-linear systems this scheme allows to stabilize the unstable
orbits for which the condition (\ref{q_50}) does not fulfill. The
main idea of the method \cite{Socolar:1994_ControllingChaos} is
the following: the feedback signal is formed so that not only the
system state at the moment $(t-T_k)$ influences upon it, as it was
in the previous scheme, but also the states at the moments of time
$(t-mT_k)$ do, with some weight coefficients. Following work
\cite{Franceschini:1999}, we assume that the described scheme
could be effective in the controlling dynamics of the distributed
chaotic system.

The continuous feedback for stabilization of the unstable periodic
states of higher period $T_k>T_1$ can be described by the
following expression $$ \varphi(x=1.0,t)=K\xi(t)=$$
\begin{equation}\label{q_OSS_1}
 =K\left((1-r)\sum\limits_{m=1}^Mr^{m-1}
  \left(\rho(x_{\rm fix},t)-\rho(x_{\rm fix},t-md)\right)\right),
\end{equation}
where the delay time $d$ is taken equal to the period of the
stabilized periodic orbit $T_k$, $M$ is large enough ($M\gg1$),
and $r$ ($0\leq r<1$) characterizes the contribution of the
previous system's states into the feedback signal: small values of
$r$ correspond to the small significance of the previous states in
the signal (\ref{q_OSS_1}), and analogously, correspond to the
large ones. The case $r=0$ corresponds to the simplest scheme
(\ref{q_OSS}) of continuous feedback discussed above. Such
feedback scheme (\ref{q_OSS_1}) can be realized in practice, for
example, by using a set of acoustic time delay lines each with its
own delay time and transfer constant. As it follows from works
\cite{Franceschini:1999, Bleich:1999_ControlChaos,
Simmendinger:1999_ControlChaos}, a possibility of stabilization of
the orbits of highest period $T_k$ (\ref{q_50}) appears with the
growth of $r$, for which the use of the scheme (\ref{q_OSS}) is
impossible. For example, in the work \cite{Just:1999_book} it is
shown that with the help of the scheme similar to (\ref{q_OSS_1})
it is possible to stabilize the unstable periodic states for which
the following estimation condition is fulfilled:
\begin{equation}\label{q_50a}
 \Lambda T\leq C\frac{1+r}{1-r}.
\end{equation}

Using the scheme (\ref{q_OSS_1}) for the stabilization of the
periodic orbits which fulfills the condition (\ref{q_50a}), the
explicit knowledge of the orbit period $T_k$ is needed. If the
delay time $d$ differs even slightly from the real period $T_k$,
it is impossible to stabilize the unstable periodic state of the
system, and in the feedback line one can observe periodic
oscillations $\xi(t)$ with the amplitude exceeding significantly
the noise level in the system. As numerical analysis shows,
stabilization is possible only when the orbit period is known with
the precision $\varepsilon<0.01\%$.

For more accurate definition of the orbit periods $T_k$ the
following method presented in work \cite{Kittel:1995_ControlChaos}
for the systems with few degrees of freedom and approbated for
distributed systems in the work \cite{Franceschini:1999} is used.

If the delay time $d$ is not explicitly ``tuned'' to the orbit
period $T_k$ then the maximum Lyapunov exponent is negative and in
the feedback line one can observe the oscillations with the
amplitude exceeding significantly the noise level of the system
and the ``base'' period $\Theta\not= T_k$, which depends upon $K$
and $d$ (see. Fig.\,\ref{ref14}{\it d,e}). In work
\cite{Just:1999_book} the analytical dependence between the
unknown explicit period $T_k$ of the unstable state, feedback
parameters $K$, $d$ and the period $\Theta$ has been found as
follows:
\begin{equation}\label{q_51}
 \Theta(K,d)=T_k+\frac{K}{K-g}(d-T_k)+{\rm O}({(d-T)^2}).
\end{equation}
Here $g$ is the unknown parameter which is defined by the type of
the non-linear dynamical system and depends on the form of the
delayed feedback. Thus, in expression (\ref{q_51}) there are two
unknown values: the period $T_k$ of the unstable periodic state
and the parameter $g$.

Let us take two sets of values $(K_1,d_1)$, $(K_2,d_2)$, where the
values of $d_{1,2}$ are chosen close to the period of the
stabilized orbit, and then find the corresponding $\Theta_1$ and
$\Theta_2$. Then we find numerically the solution of the system of
two non-linear equations (\ref{q_51}) and determine the values $g$
and $T_k$. Then we repeat the procedure taking the newly defined
$T_k$ as $d$. After a number of such iterations the period of the
unstable orbit $T_k$, would be defined with absolutely
satisfactory precision and then the scheme (\ref{q_OSS_1}) could
be used for stabilization of this state.

Numerical analysis shows that with the help of this method it is
possible to stabilize the orbits with the period $T<25.0$. The
unstable states with higher period turned out to be impossible to
stabilize, though in the system the periodic oscillations can be
observed, but the form of these oscillations is not close to the
concerned unstable state and the feedback signal is not small.
With the growth of the period $T$ of the stabilized orbit it was
necessary to enlarge the parameters $M$ and $r$ to achieve the
effect of controlling chaos. Simultaneously, the range of value of
the feedback coefficient $K$, for which stabilization of the
unstable state is possible, narrows.

\begin{figure*}
\centerline{\scalebox{0.65}{\includegraphics{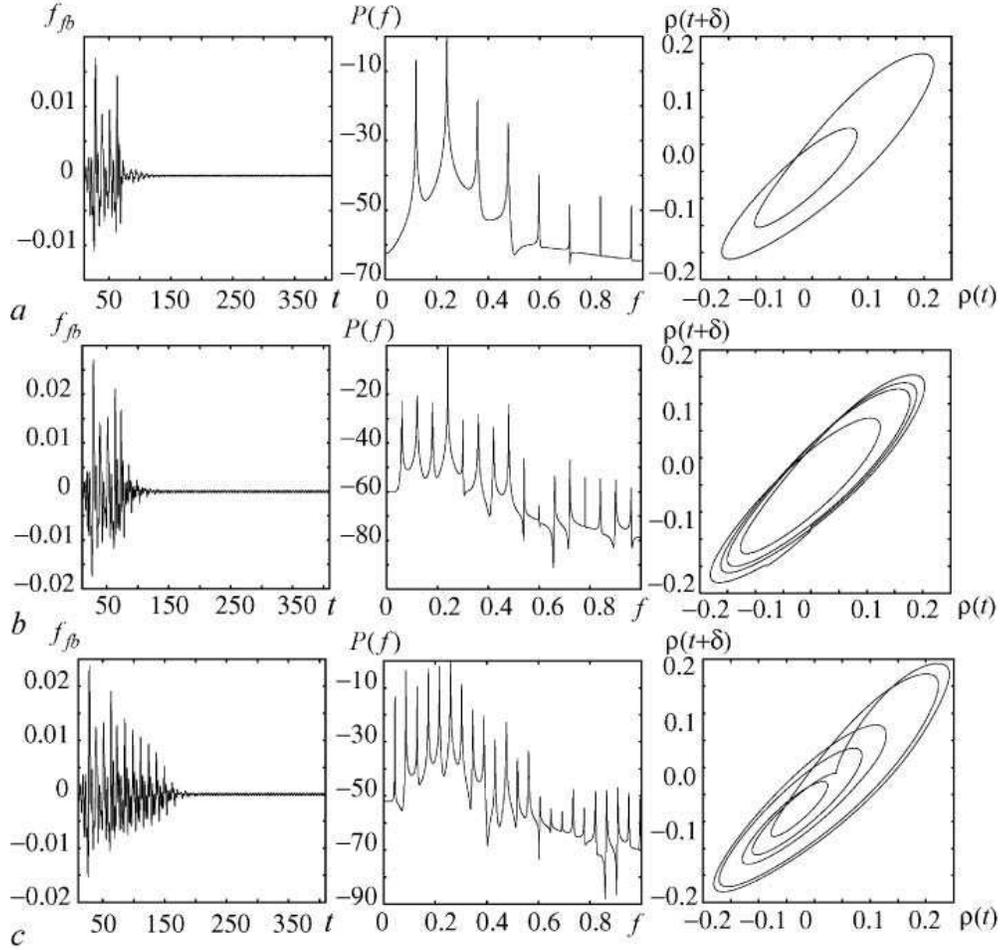}}}
\caption{Stabilization of the unstable states of high period:
({\it a}) $T=8.3829$ ($M=3$, $K=0.03$, $r=0.2$); ({\it b})
$T=16.6582$ ($M=12$, $K=0.012$, $r=0.38$); ({\it c}) $T=23.1286$
($M=20$, $K=0.008$, $r=0.56$). The pictures (from left to right)
present: the time series of feedback signal $f_{\rm fb}$, the
power spectrum of the oscillations in stabilized system and the
phase portrait for the space charge density oscillations
$\rho(x_{\rm fix}=0.2,t)$. The feedback is switched on at the
moment of time $t=50.0$. Delay time of time-delayed embedding
$\delta=1$}
 \label{ref15}
\end{figure*}

Stabilization of the orbits of highest period is illustrated by
Fig.~\ref{ref15}. The feedback scheme parameters $M$, $r$ and $K$
are defined in the cutline. The explicit values of the periods $T$
are defined according to the method described above. In
Fig.~\ref{ref15} the orbits of the period $T\approx8$,
$T\approx16$ and $T\approx23$ are shown. The latter is the
unstable orbit with the highest period which appears to be
possible to stabilize in spiral chaos regime ($\alpha=2.857\pi$)
with the help of the discussed scheme.

\section*{Conclusion}

In this work the method of controlling complex chaotic dynamics of
the spatially distributed active medium ``electron beam with
overcritical current in Pierce diode'' is discussed. The described
method is based on the ideas of controlling chaos in non-linear
systems with few degrees of freedom. The scheme of continuous
delayed feedback, which is used for controlling, allows to
stabilize the unstable equilibrium state of the distributed system
and the unstable periodic spatio--temporal states analogous to the
unstable periodic orbits of the chaotic attractor in the systems
with few degrees of freedom.

\section*{Acknowledgements}

The authors are thankful to Professor D.I.~Trubetskov for the
interest to this work. We also thank Dr. Svetlana V. Eremina for
English language support. This work has been supported by the
Russian Foundation for Basic Research (grant 05--02--16286).
A.E.H. and A.A.K. also thank ``Dynasty'' Foundation and ICFPM for
financial support. A.E.H. acknowledges support from CRDF, Grant
No.~Y2--P--06--06.


\end{document}